\documentclass[twocolumn,
longbibliography,
prb]{revtex4-1} 
\usepackage{graphicx} 
\usepackage{graphicx,multirow,amssymb,amsmath,fullpage} 
\usepackage{dcolumn,multirow,verbatim} 
\usepackage{amsmath} 
\usepackage{bm} 
\usepackage[margin=0.75in]{geometry} 
\usepackage{wrapfig}
\usepackage{url}
\usepackage{hyperref}
\hypersetup{colorlinks,breaklinks,linkcolor=Blue,urlcolor=Blue,anchorcolor=Blue,citecolor=Blue}
\usepackage{color}
\definecolor{DarkBlue}{rgb}{0.0,0.08,0.45}
\definecolor{Blue}{rgb}{0.0,0.0,1.0}
\definecolor{Red}{rgb}{1.0,0.0,0.0}
\definecolor{RedOrange}{rgb}{0.9,0.0,0.2}
\definecolor{dgrn}{RGB}{0,150,0}
\definecolor{dgray}{gray}{0.3}
\newcommand*{\citen}{}
\DeclareRobustCommand*{\citen}[1]{%
  \begingroup
    \romannumeral-`\x 
    \setcitestyle{numbers}%
    \cite{#1}%
  \endgroup
}

\newcolumntype{.}{D{.}{.}{-1}}
\newcommand{\mc}[3]{\multicolumn{#1}{#2}{#3}}

\begin{document}

\title{
On the Spin Symmetry of Fractional-Occupation Mean-Field Approaches
}
\author{Md Rafi Ul Azam}
\author{Rachel Campbell}
\author{Kay Carter-Fenk$^\ast$}
\affiliation{Department of Chemistry, University of Pittsburgh, Pittsburgh, Pennsylvania 15260, USA\\
\href{mailto:kay.carter-fenk@pitt.edu}{\texttt{kay.carter-fenk@pitt.edu}}
}
\date{\today} 
\begin{abstract}
We identify a spin-symmetry dilemma rooted in the representability problem of fractional-occupation densities, which break spin symmetry even for restricted orbital references. We propose a resolution that leverages complex polarization to restore pure-state $N$-representability and exact spin eigenstates while retaining improvements to the energy and density from occupation smearing. Our results have implications for widely-used smearing methods in solid-state simulation and for their use as a proxy for static correlation in molecules.
\end{abstract}

\maketitle

Mean-field electronic structure methods, including Hartree-Fock (HF) theory and Kohn-Sham density-functional theory (KS-DFT), owe much of their success to their compact single-determinant -- or pure-state $N$-representable -- description of the electronic-structure problem.\cite{HohKoh64,KohSha65,SzaOst82,ParYan89}  However, when multiple configurations contribute to the ground-state wave function, static correlation arises and the single-determinant picture breaks down. This scenario is encountered across many problems in molecular physics including bond dissociation,\cite{GriSchBae97, FucNiqGon05} transition states,\cite{ChaSim18} diradicals,\cite{SalRow72} and in bimetallic transition-metal complexes.\cite{CohMorYan08b, SheLoiHai21}

A common calling card of static correlation in single-determinant models is the presence of lower-energy spin-symmetry-broken solutions.
Such is the classic symmetry dilemma: \cite{LykPra63,PerSavBur95} broken-symmetry solutions may give more reasonable energetics, but they are not eigenfunctions of the total-spin operator, $\hat{S}^2$.\cite{Low55,BauAhl96b} Spin-symmetry breaking can therefore indicate qualitative deficiencies in the wave function,\cite{SheLoiHai21} with measurable consequences for solid-state observables such as magnetic dipolar couplings and zero-field splittings.\cite{BikSchGer20} These deficiencies can be remedied, at least approximately, through spin projection\cite{Low55,Viv19,Sch88,Sch98, JimHenTsu12,FerGuiMal15,HenScu17,KitSaiYam18,DavTriMal20} to improve calculated properties while restoring essential symmetries of the wave function.

Configuration interaction (CI) is a straightforward approach to mend spin-symmetry breaking in the reference wave function while simultaneously addressing static correlation.\cite{Low55,HenScu17} However, its exponential cost has motivated polynomial-scaling models that approximately account for multi-configurational effects without explicit access to the wave function. 
In particular, fractional orbital occupations offer an affordable route to capturing static correlation\cite{MarShaJia23} within an independent-particle framework that arises naturally in ensemble descriptions,\cite{Col63,PerParLev82,GroOliKoh88b, Nes97} finite-temperature formalisms, \cite{Mer65} reduced density-matrix methods,\cite{Gil75,McW92, YanZhaAye00, GibBoyMaz22b} and natural-orbital functional theory.\cite{Pir17,Pir18b}
Fractional charges and fractional spins have also been employed to diagnose and correct static correlation error and spin-state energetics in DFT.\cite{WanSch96,CohMorYan08a,HesJohHu11,YanMorCoh13,SuLiYan18}

Beyond acting as a proxy for static-correlation, fractional charges are frequently employed in solid-state calculations as a numerical aid to self-consistent field (SCF) convergence near band crossings\cite{VerGon01, SanMar23} and separately as a physical model of finite-temperature electronic structure.\cite{Mer65, PitProFlo11} The computational solid-state physics community relies on fractional occupation models,\cite{MetPax89,MarVanDeV99} but there has been little study into the impacts that fractional charge schemes have on spin symmetry. Spin symmetry governs selection rules for spin-dependent phenomena in magnetic materials\cite{SivDosIba25,ZhaYeYao24} and for optical excitations.\cite{EchUrbAma16,LiDerDer24} Moreover, many practical spin-orbit coupling workflows construct a spin-orbit Hamiltonian on top of a scalar-relativistic reference,\cite{Nee05, RoySan17, ZhaCha22} making the spin character of that reference directly consequential for any property derived from the subsequent spin--orbit treatment. 
It is therefore essential that we understand the presence and nature of any artificial spin-symmetry breaking that fractional-charge models might introduce.

Though our findings here are general, we will focus on thermally-assisted occupation (TAO) mean-field methods that use Fermi--Dirac smearing with a fictitious electronic temperature, $\theta$, to encode static correlation into the density at $\mathcal{O}(N^3)$ cost.\cite{Cha12,Cha14,WuCha15,YehCha16,LinHuiChu17,Cha17,CheCha22,GibBoyMaz23}
Recent work on TAO models suggests that a sufficiently large $\theta$ can restore spatial spin-density symmetry in unrestricted TAO-DFT.\cite{WanCha24} However, fractional occupations fundamentally alter the algebraic structure of the one-particle density matrix, resulting in a state that is ensemble $N$-representable but not pure-state $N$-representable. In this letter, we identify a spin-symmetry dilemma that manifests ubiquitously in fractional-occupation schemes due to this representability problem and we propose a simple resolution.


Specifically, we derive $\langle S^2 \rangle$ analytically without the usual assumption of an idempotent density matrix for restricted, unrestricted, and generalized Hartree-Fock reference states. We use these results to demonstrate that standard Fermi--Dirac smearing severely degrades $\langle S^2 \rangle$ even as it reduces spatial spin polarization, showing that spin-density symmetry restoration and spin-angular-momentum consistency are distinct properties in the fractional-occupation construction. We then show that both can be reconciled by expressing the fractional-occupation density matrix in a complex Hermitian form.\cite{LeeBerSma19}
We demonstrate that such a framework results in total recovery of spin symmetry in single- and multiple-bond dissociation using restricted spin-orbitals. We also show an improved agreement between our purified fractional-occupation densities and those from complete active-space self-consistent field (CASSCF) theory.  Our results establish a density matrix route to resolving the spin-symmetry dilemma in fractional-occupation models while retaining their useful correlated-density description, implying the existence of a principled path to incorporate static correlation into (single-configuration) KS-DFT.

Without the usual assumption of an idempotent density matrix, the total-spin expectation value evaluated in the atomic orbital basis is,
\begin{equation}
\begin{aligned}
\langle \hat{S}^2 \rangle_{\mathrm{UHF}}
={}&
\frac{1}{4}(N_{\alpha}-N_{\beta})^2
\\
&\hspace{-2.8em}
+\frac{1}{4}\Big[
\left(N_{\alpha}-\mathrm{Tr}[\mathbf{P}_{\alpha}\mathbf{S}\mathbf{P}_{\alpha}\mathbf{S}]\right)
+\left(N_{\beta}-\mathrm{Tr}[\mathbf{P}_{\beta}\mathbf{S}\mathbf{P}_{\beta}\mathbf{S}]\right)
\Big]
\\
&+\frac{1}{2}(N_{\alpha}+N_{\beta})
-\mathrm{Tr}[\mathbf{P}_{\alpha}\mathbf{S}\mathbf{P}_{\beta}\mathbf{S}] .
\label{eq:s2_uhf}
\end{aligned}
\end{equation}
where $\mathbf{S}$ is the atomic-orbital overlap matrix, $N_\alpha$ and $N_\beta$ are the number of $\alpha$- and $\beta$-spin electrons, and $\mathbf{P}_\sigma$ is the $\sigma=\alpha,\beta$ density matrix.
In a restricted orbital representation with idempotent $\mathbf{P}_\sigma$, the idempotency residual (term in square brackets) vanishes, and the overlap trace (last term) reduces to $N_\beta$, yielding $\langle S^2\rangle=S(S+1)$, the exact result. We note that the above expression holds for any non-interacting $\mathbf{P}_\sigma$ (ensemble or pure-state), so while much of our discussion is centered on TAO approaches, our findings can be applied to other fractional-occupation models.

In contrast, the idempotency residual does not vanish for a non-interacting fractional-occupation density matrix and the overlap trace is no longer guaranteed to equal $N_\beta$, even when $\mathbf{P}_\alpha=\mathbf{P}_\beta$. Thus, occupation smearing generates nonzero $\langle S^2\rangle$ even within a restricted orbital representation. Notably, this counterintuitive feature is a generalization of the fractional-spin H-atom result reported by Loos and co-workers\cite{BurMarDaa21} to many-electron systems. The resultant spin-symmetry breaking in the restricted case is a direct consequence of non-idempotency that manifests due to a lack of pure-state $N$-representability and is not necessarily related to spatial spin polarization. We denote this as a unique spin-symmetry dilemma: incorporating static correlation into the density via fractional-charge approaches can rectify qualitative failures of single-configuration densities, but at the cost of spin symmetry.

We now explore a resolution to this dilemma that allows the recovery of spin-symmetry while retaining a description of static correlation at the mean-field level. In particular, restoring pure-state $N$-representability via idempotency constraints would ameliorate both the non-vanishing idempotency residual and the inequality between $N_\beta$ and the overlap trace.
To reconcile fractional occupations with a well-defined $\langle S^2 \rangle$, we use the complex-density representation of Head-Gordon and coworkers as a guidepost,\cite{LeeBerSma19}
\begin{equation}
\mathbf{P}_\sigma = \mathbf{X}_\sigma + i\mathbf{Y}_\sigma\;,
\label{eq:complex_density}
\end{equation}
in which a real symmetric density can be augmented with an antisymmetric imaginary component to achieve a Hermitian complex projector.
Whereas this additional variational degree-of-freedom was previously used to identify real-to-complex instabilities, we leverage the added flexibility for a different purpose. We initialize $\mathbf{X_\sigma}$ from the Fermi--Dirac-smeared reference, and find the nearest $\mathbf{Y_\sigma}$ that recovers pure-state $N$-representability. 
In an orthonormal representation, idempotency requires
\begin{align}
\mathbf{X}_{\sigma}^2-\mathbf{Y}_{\sigma}^2 &= \mathbf{X}_{\sigma}
\label{eq:x_constraint}
\\
\mathbf{X_{\sigma}Y_{\sigma}+Y_{\sigma}X_{\sigma}} &= \mathbf{Y}_{\sigma}
\label{eq:y_constraint}
\end{align}
These constraints, therefore, define the conditions under which idempotency may be restored without density-matrix purification collapsing the fractional reference into an integer-occupation density.


 We use this foundation to purify the density matrix by optimizing an objective function that minimizes the change in the occupation spectrum in a least-squares sense (details provided in the Appendix). This protocol results in the automated selection of occupied--virtual pairs that satisfy equations~\ref{eq:x_constraint} and \ref{eq:y_constraint} while minimally disrupting the occupation spectrum.
Once paired, the elements of $\mathbf{X}_\sigma$ satisfy $n_{i\sigma}+n_{a\sigma}=1$ and the corresponding elements of $\mathbf{Y}$ are,
\begin{equation}\label{eq:Ypart}
Y_{ia,\sigma}=\sqrt{n_{i\sigma}n_{a\sigma}},
\end{equation}
for all perfect pairs, $i\leftrightarrow a$.
For a single pair, the corresponding block of the idempotent density matrix is
\begin{equation}
\mathbf{P}_{ia,\sigma}
=
\begin{pmatrix}\label{eq:twobytwo}
X_{ii,\sigma} & -iY_{ia,\sigma}\\
iY_{ia,\sigma} & X_{aa,\sigma}
\end{pmatrix}.
\end{equation}
While energies are not the focal point of this work, we account for the energy change due to static correlation using the deviation from idempotency of the real part of the density matrix via $\gamma \text{Tr}[\mathbf{X}^2-\mathbf{X}]$,\cite{GibBoyMaz23} where $\gamma$ is a parameter.

Complex density-matrix construction has been studied before, but this is the first time that the commutation relations in equations~\ref{eq:x_constraint} and \ref{eq:y_constraint} have been used to ameliorate spin contamination in fractional-occupation models.
We next identify the presence of this heretofore unidentified symmetry dilemma in fractional-occupation approaches with a few simple examples and show that our complex construction removes artificial spin-symmetry breaking caused by fractional charges.
Our results demonstrate that such lossy information-compression algorithms can be used in the rank reduction of ensemble density matrices, preserving their static-correlation content but restoring a Kohn-Sham single-determinant construction. Throughout our analysis, we use HF wave functions as to sidestep concerns pertaining to the definition of spin eigenfunctions in the KS-DFT context.\cite{JacRei12}

\begin{figure}[t]
\includegraphics[width=\columnwidth]{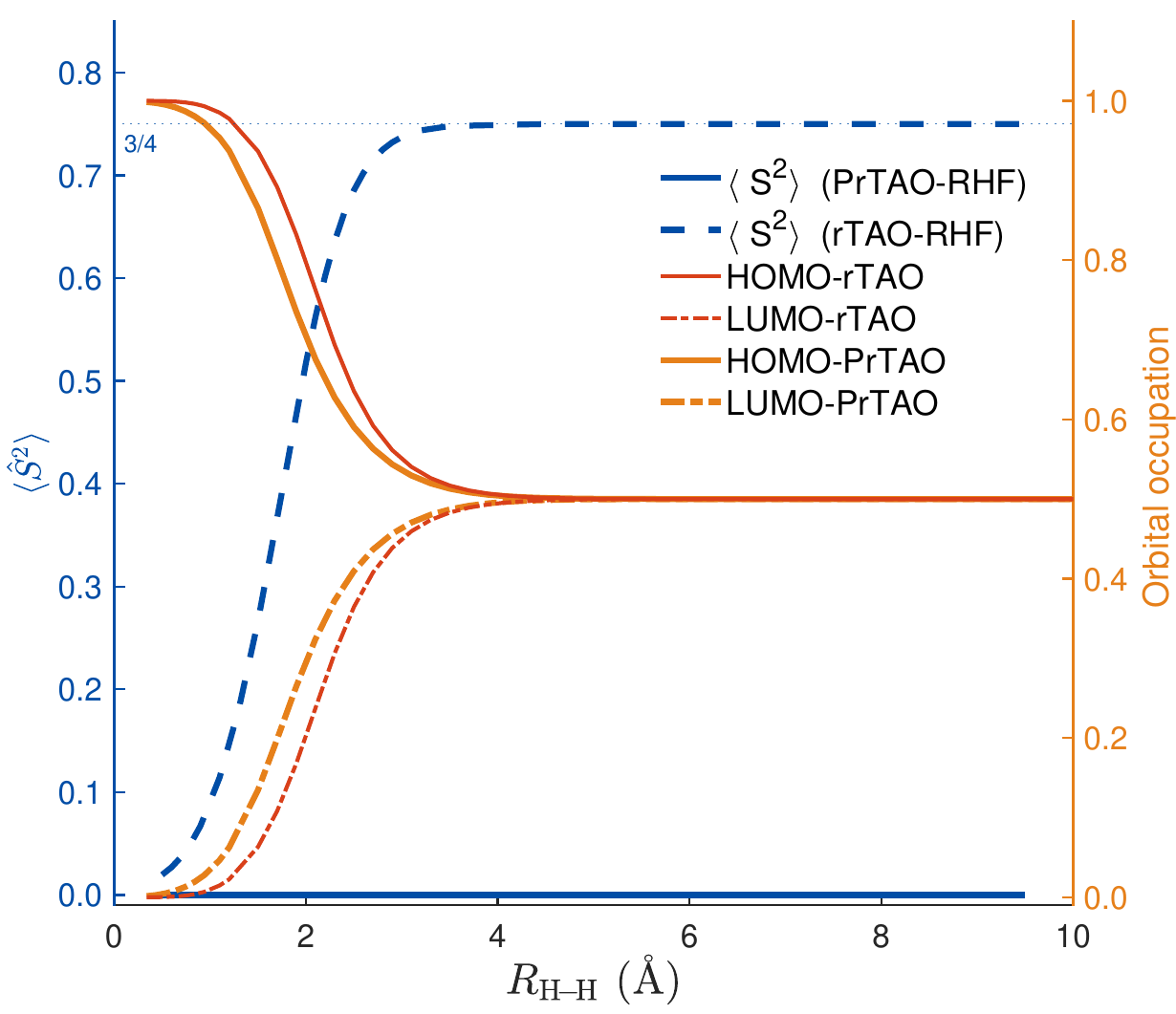}
\caption{Fractional occupations and $\langle S^2\rangle$ for 
${\text{H}_2}$ dissociation computed using restricted orbitals in a STO-3G basis with rTAO-HF(2,2) and our PrTAO-HF(2,2) procedure at $\theta = 100~$m$\text{E}_\text{H}$.}
\label{fig:h2}
\end{figure}

First, we study the bond dissociation of $\text{H}_2$ molecule using a 2-electron, 2-orbital (2,2) active space for Fermi--Dirac smearing at the HF/STO-3G level. The results in Fig.~\ref{fig:h2} show that the conventional rTAO begins to break spin symmetry almost immediately, reaching an asymptotic $\langle S^2\rangle=0.75$. This result is sensible, as it is precisely twice the $\langle S^2\rangle=3/8$ for half-occupied orbitals in the fractional-spin H atom.\cite{BurMarDaa21} Turning our attention to the complex-polarized result in our purified rTAO (PrTAO) model, we see a smooth fractionalization of occupation numbers to the  correct asymptotic limit of half-occupied highest-occupied and lowest-unoccupied molecular orbitals (HOMO and LUMO, respectively). However, our analytic purification procedure makes these fractional-occupied densities pure-state $N$-representable at every point on the dissociation curve, allowing for exact recovery of $\langle S^2\rangle=0$ everywhere. 


\begin{figure}[t]
\includegraphics[width=\columnwidth]{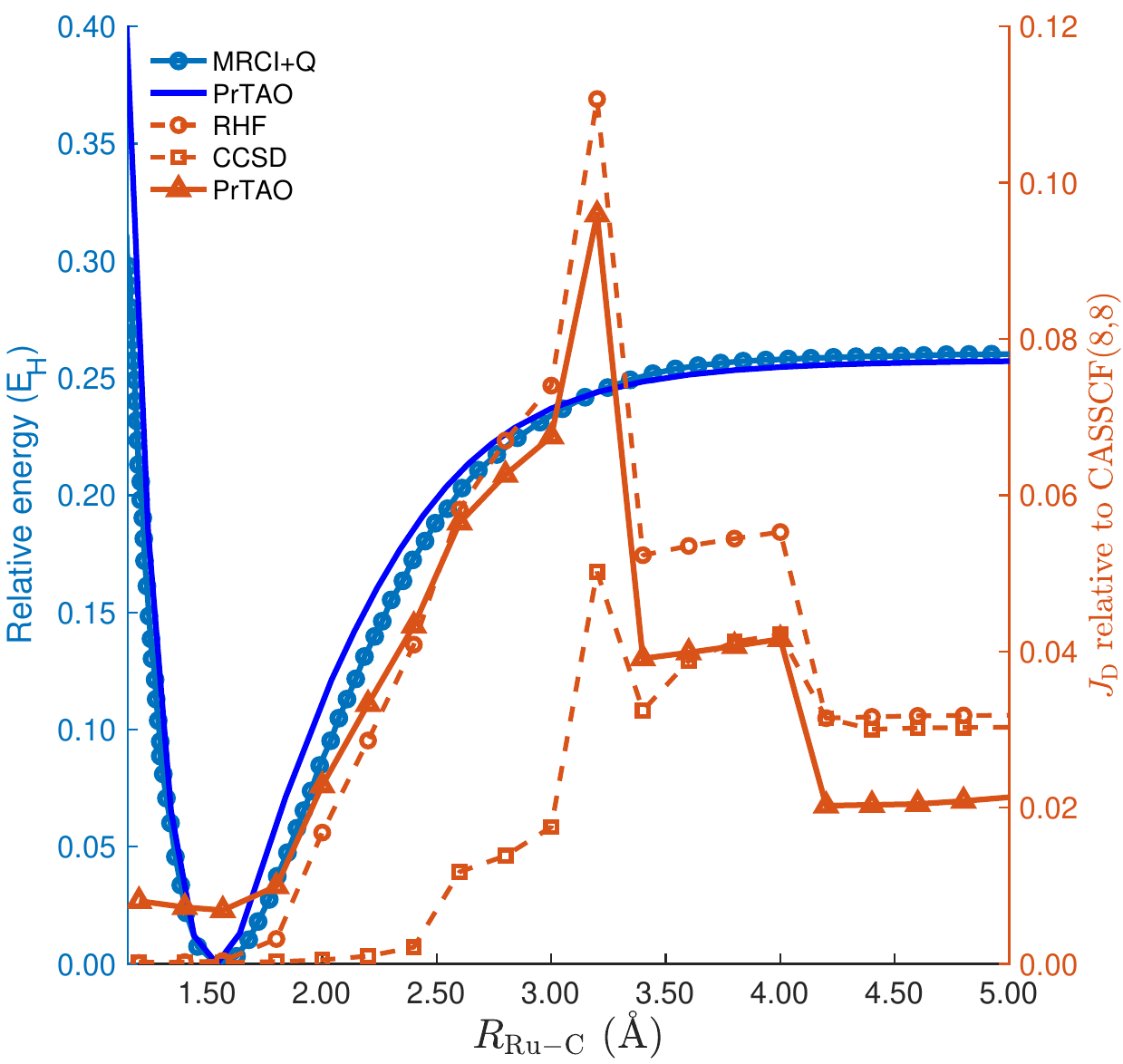}
\caption{Relative potential-energy curves for RuC dissociation using PrTAO-HF(8,8) ($\theta=50$~m$\text{E}_{\text{H}}$ and $\gamma = 2$~$\rm E_H$)and Jeffreys divergences relative to the CASSCF(8,8) density using restricted reference orbitals. Canonical HF and CCSD Jeffreys divergences ($J_D$) are reported for comparison. Single-point energies were evaluated using aug-cc-pV5Z\cite{Dun89} for C with the corresponding -PP basis for Ru, and $J_D$ were evaluated with Def2-TZVPP.\cite{def2}}
\label{fig:RuC}
\end{figure}


In Fig.~\ref{fig:RuC} we examine the asymmetric quadruple-bond dissociation curve of singlet RuC. While the bond order and the asymmetry of the problem make it significantly more challenging than H$_2$,  PrTAO-HF(8,8) is capable of semi-quantitatively reproducing the multi-reference configuration interaction with quadruples corrections (MRCI+Q) relative potential energy surface.\cite{TzeKar20} Over the extent of the bond dissociation curve, we note that $\langle S^2\rangle=0$ for our PrTAO scheme, whereas rTAO with restricted orbitals asymptotically approaches $\langle S^2\rangle=2.9$, exhibiting catastrophic symmetry breaking.

\begin{table*}[t]
\centering
\caption{Jeffreys divergences ($\times10^{-3}$) of the total density $n$ relative to CASSCF benchmarks for closed-shell, collinear systems and GCCSD for Cr$_3$. For Cr$_3$, $|\mathbf M|^2$ is the magnetization length. The same active space is used for both CAS and PrTAO. All calculations are done using def2-TZVPP, \cite{def2} except O$_3$ and Be$_2$ use cc-pVTZ.\cite{PetDun02}}
\label{tab:jeffreys}

\begin{tabular}{l c c c c c c c | c c}
\hline\hline
 & \mc{7}{c|}{Restricted} & \mc{2}{c}{Generalized}\\
\cline{2-8}  \cline{9-10}
\multirow{2}{*}{System} & \multirow{2}{*}{O$_3$$^a$}
& \multirow{2}{*}{BN$^b$}
& \multirow{2}{*}{Be$_2$$^c$}
& \multirow{2}{*}{C$_2$$^d$}
& \multirow{2}{*}{CH$_2$$^e$}
& \multirow{2}{*}{F$_2$$^f$}
& \multirow{2}{*}{SiH$_2$$^g$}
& \mc{2}{c}{Cr$_3$$^h$}\\

& & & & & & & & $n$ & $|\mathbf{M}|^2$\\
\cline{9-10}

\hline

\mc{1}{l}{HF}
& 3.6
& 17.2
& 2.7
& 3.3
& 1.0
& 0.6
& 0.8
& 0.7 
& 84\\

\mc{1}{l}{CCSD}
& 1.6
& 0.9
& 1.1
& 0.2
& 0.3
& 0.1
& 0.1
& -- & -- \\


\mc{1}{l}{PrTAO-HF}
& 1.2
& 11.9
& 2.3
& 3.1
& 0.9
& 0.5
& 0.7
& 0.7 
& 78\\

\hline\hline
\mc{10}{l}{\footnotesize $^a\theta=430~{\rm mE_H}$, active space: (6,6)}\\
\mc{10}{l}{\footnotesize $^b\theta=70~{\rm mE_H}$, active space: (8,8)}\\
\mc{10}{l}{\footnotesize $^c\theta=30~{\rm mE_H}$, active space: (8,8)}\\
\mc{10}{l}{\footnotesize $^d\theta=40~{\rm mE_H}$, active space: (8,8)}\\
\mc{10}{l}{\footnotesize $^e\theta=50~{\rm mE_H}$, active space: (4,4)}\\
\mc{10}{l}{\footnotesize $^f\theta=80~{\rm mE_H}$, active space: (14,14)}\\
\mc{10}{l}{\footnotesize $^g\theta=40~{\rm mE_H}$, active space: (8,8)}\\
\mc{10}{l}{\footnotesize $^h\theta=40~{\rm mE_H}$, active space: (16,16)}
\end{tabular}
\end{table*}

We also appraise the quality of the PrTAO-HF(8,8) density using the Jeffreys divergence ($J_{\mathrm D}$).\cite{Jef46, ZamCar26} 
In Fig.~\ref{fig:RuC}, we compare RHF, CCSD, and PrTAO-HF(8,8) densities to CASSCF(8,8), noting that smaller values of $J_{\mathrm D}$ indicate greater similarity between the test and reference densities.
As the bond dissociates and static correlation becomes dominant, the PrTAO-HF(8,8) begins to improve upon RHF, matching the quality of the CCSD density at around 3.4~\AA. While fractional-occupation models rooted in Fermi--Dirac smearing cannot boast close correspondence with CASSCF occupation numbers,\cite{HuSheWan24} it is notable that the PrTAO-HF(8,8) density is much more similar to CASSCF(8,8) at dissociation than RHF or CCSD. A pronounced nonmonotonic feature in $J_{\mathrm D}$ at 3.1~\AA\ and two subsequent step features coincide with a reorganization of the frontier orbitals; their respective orbital origins are examined in the Appendix~\ref{app:ruc_orbitals}.

To study static correlation outside of a bond-breaking context, we also examined the ethylene torsional profile. Ethylene torsion forms a diradical at an angle of 90$^\circ$, where the HOMO and LUMO become exactly degenerate. Unlike the bond breaking cases examined above, the torsional profile is not recovered much more accurately by choosing unrestricted reference orbitals.\cite{ShaHeaKry03} Appendix~\ref{app:ethylene} shows that PrTAO-HF(12,12) quantitatively recovers the torsional profile of ethylene without sacrificing $\langle S^2\rangle=0$.

Motivated by the apparent improvement of the density in RuC bond dissociation, we turn our attention to how well PrTAO-HF can reproduce the densities of a series of molecules that exhibit static correlation at their equilibrium geometries. 
In Table~\ref{tab:jeffreys}, we use $J_D$ to facilitate density comparisons for 7 molecules exhibiting collinear singlet ground states along with the equilateral-triangle geometry of Cr$_3$, which exhibits a strongly-correlated non-collinear ground state.\cite{AntLazUdv08}
For all collinear systems, we use a CASSCF reference with the same active space as PrTAO-HF. We choose a two-component generalized coupled-cluster with single and double substitutions (GCCSD) reference for Cr$_3$ due to system size and the completeness of the active space required to recover the correct non-collinear ground state.

Across the collinear singlets, PrTAO-HF lowers $J_{\mathrm D}$ relative to RHF in every case. This improvement, which manifests from occupation smearing, is retained because our purification minimally perturbs the Fermi--Dirac occupation spectrum.
For the noncollinear case of Cr$_3$, we note no improvement in the total density relative to generalized HF (GHF). However, we do see some improvement in the length of the magnetization vector, suggesting that spin is somewhat better described in the PrTAO-HF framework.

Interestingly, this suggests that the two-component framework changes the role and nature of density corrections achieved via fractionalization. In GHF, fractional occupations smear over charge and spin coordinates simultaneously, so while GHF spin-symmetry breaking reduces the impact of fractionalization on the charge density, redistributing spin populations can improve the net magnetization.  Overall, our density purification sidesteps the additional spin-contamination encountered in typical rTAO-HF, giving $\langle S^2\rangle = 0$ for all collinear singlet references. We note that PrTAO-HF retains the same $\langle S^2\rangle = 8.2$ as the noncollinear GHF reference while improving the length of the magnetization vector. This echos the results of Ref.~\citen{WanCha24} while again emphasizing the nuance that improvements to spatial spin-density distributions do not necessarily carry over to improvements in spin eigenvalues.

In contrast, our analysis shows that the recovery of equal spin densities is not sufficient to guarantee recovery of singlet spin eigenstates in mean-field fractional-occupation models.
This brings us to conclude that this is a novel manifestation of a spin-symmetry dilemma in fractional-occupation methods. Although mean-field energies and densities may be improved by incorporating fractional-occupation smearing alongside corrections to the energy functional, the ensemble $N$-representable density breaks spin symmetry -- even with restricted orbitals.
By allowing for complex polarization, we have suggested a simple analytic purification procedure to restore pure-state $N$-representability (and therefore spin symmetry) while minimally perturbing the occupation spectrum.
Our work therefore clarifies the representability requirement underlying spin-adapted fractional-occupation theories and provides a mean-field route to incorporate static correlation without sacrificing spin symmetry. The results reported here can be used to guide the design of static-correlation-aware mean-field methods that correctly retain the expected symmetries of the non-relativistic Hamiltonian.

\subsection*{End Matter}
\appendix
\section{Occupation Pairing and Cost-Matrix Construction}
We define a cost matrix to resolve the orbital-pairing ambiguity that arises within near-degenerate PrTAO active spaces. For each spin channel $\sigma$, canonical occupied and virtual orbitals are paired by their occupations according to the complementary-pair condition,
\begin{equation}
n_{i}+n_{a}=1.
\label{eq:pair_complement}
\end{equation}
Fermi--Dirac occupations are dominated by pairs $f_i\leftrightarrow f_a$ that already closely resemble a perfect-pairing occupation scheme.
To minimally perturb the occupation spectrum through purification, we
optimize,
\begin{equation}
    f(n_i)=(n_i-f_i)^2+(n_a-f_a)^2\;~.
\end{equation}
This implies that the closest occupation numbers to the original Fermi--Dirac ones that satisfy perfect pairing are,
\begin{subequations}\label{eq:occ_shift}
    \begin{equation}
        n_i = \frac{1+f_i-f_a}{2}
    \end{equation}
    \begin{equation}
        n_a=\frac{1-f_i+f_a}{2}
    \end{equation}
\end{subequations}

To minimize the average shift across the full spectrum of occupation numbers, we minimize,
\begin{equation}
C_{ia}^{\sigma} = \left| f_{i}+f_{a(i)}-1 \right|,
\label{eq:cost_matrix}
\end{equation}
such that orbitals whose Fermi--Dirac occupations are already complementary incur zero cost. The index $a(i)$ corresponds to canonical virtual $a$ paired with $i$.
The assignment with the smallest total cost is chosen. Once a partner $a(i)$ is assigned to orbital $i$,
the original Fermi--Dirac occupations are shifted according to Eq.~\ref{eq:occ_shift}.
Thus, the occupation correction is distributed symmetrically between the two members of the pair rather than being assigned entirely to one orbital.
For each complementary pair, the corresponding $2\times2$ block of the density matrix is constructed according to equations~\ref{eq:Ypart} and \ref{eq:twobytwo}. This construction exactly satisfies the conditions required for idempotency.

\section{$\langle S^2 \rangle$ for GHF}
For a generalized Hartree--Fock (GHF) determinant, the one-particle density matrix is written in spin-block form as
\begin{equation}
\mathbf{P}
=
\begin{pmatrix}
\mathbf{P}^{\alpha\alpha} & \mathbf{P}^{\alpha\beta} \\
\mathbf{P}^{\beta\alpha}  & \mathbf{P}^{\beta\beta}
\end{pmatrix},
\end{equation}
where the off-diagonal spin blocks allow non-collinear spin mixing. The corresponding total-spin expectation value in a nonorthogonal atomic-orbital basis (without assuming idempotent $\mathbf{P}$) is,
\begin{align}
\langle \hat S^2 \rangle_{\mathrm{GHF}}
={}&
\frac{3}{4} \mathrm{Tr}\!\left[
(\mathbf{P}^{\alpha\alpha}
+\mathbf{P}^{\beta\beta})\mathbf{S}
\right]\\
&+\frac{1}{4}
\left[
\mathrm{Tr}\!\left[
(\mathbf{P}^{\alpha\alpha}
-\mathbf{P}^{\beta\beta})\mathbf{S}
\right]
\right]^2
\nonumber\\
&+
\left|
\mathrm{Tr}\!\left[
\mathbf{P}^{\alpha\beta}\mathbf{S}
\right]
\right|^2
\nonumber\\
&-\frac{1}{4}
\mathrm{Tr}\!\left[
\mathbf{P}^{\alpha\alpha}\mathbf{S}
\mathbf{P}^{\alpha\alpha}\mathbf{S}
\right]
-\frac{1}{4}
\mathrm{Tr}\!\left[
\mathbf{P}^{\beta\beta}\mathbf{S}
\mathbf{P}^{\beta\beta}\mathbf{S}
\right]
\nonumber\\
&-
\mathrm{Tr}\!\left[
\mathbf{P}^{\alpha\alpha}\mathbf{S}
\mathbf{P}^{\beta\beta}\mathbf{S}
\right]
+\frac{1}{2}
\mathrm{Tr}\!\left[
\mathbf{P}^{\alpha\beta}\mathbf{S}
\mathbf{P}^{\beta\alpha}\mathbf{S}
\right]
\label{eq:s2_ghf}
\end{align}
Note that beyond the issues discussed with non-idempotent densities in the main text, two-component generalizations of fractional-occupation models
smear both charge and spin.

\section{Twisted angle of ethylene}\label{app:ethylene}

\begin{figure}[t]
\includegraphics[width=\columnwidth]{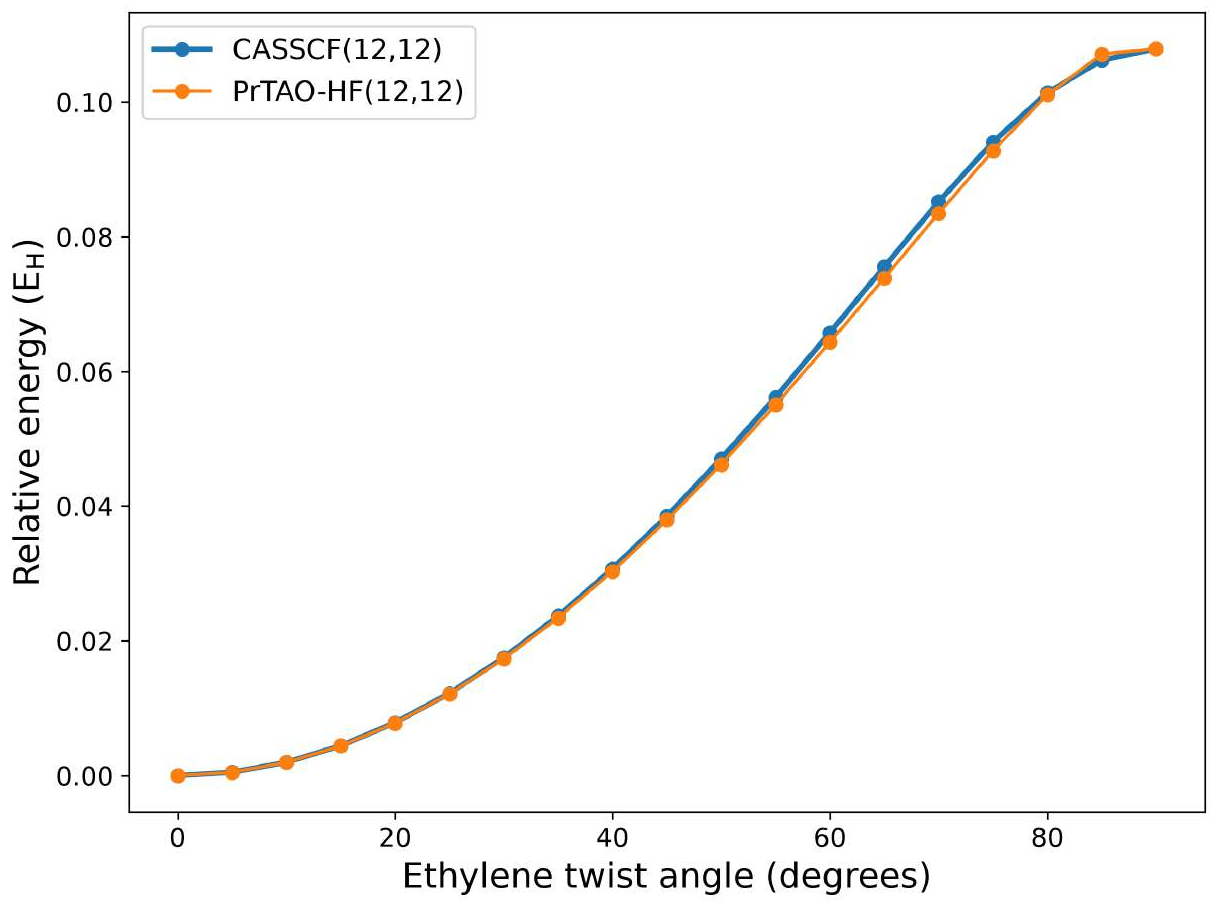}
\caption{Relative torsional potential-energy curves for ethylene computed using PrTAO-HF(12,12) and CASSCF(12,12) as a function of the twist angle at $\theta = 72$~$\rm mE_H$ and $\gamma = 5/2$~$\rm E_H$. Single-point energies were evaluated using cc-pVDZ. \cite{Dun89}}
\label{fig:C2H4}
\end{figure}

We examine torsion about the C=C bond of ethylene, where static correlation is introduced by geometric distortion rather than by nuclear separation, and where the exact singlet remains the ground state at every twist angle. We benchmark PrTAO-HF(12,12) against CASSCF(12,12). As the twist angle increases, the $\pi$ and $\pi^\ast$ orbitals become progressively closer in energy and their occupations approach the half-occupied limit as the system becomes more diradical in character.
PrTAO-HF yields a barrier nearly coincident with the CASSCF reference at the perpendicular geometry, where correlation effects are largest and a single determinant is least adequate. Semi-quantitative results are reported for the entirety of the torsional profile. All the while, our density-purification strategy maintains pure-state $N$-representability and recovers the appropriate $\langle S^2\rangle=0$ throughout the torsion. PrTAO-HF(12,12) recovers static correlation without the spin contamination accompanying conventional symmetry-broken treatments of this system.\cite{ShaHeaKry03}

\section{Frontier orbital reorganization during RuC dissociation}
\label{app:ruc_orbitals}
\begin{figure}[t]
\centering 
\begin{minipage}{0.48\columnwidth}
    \centering
    \includegraphics[width=\linewidth]{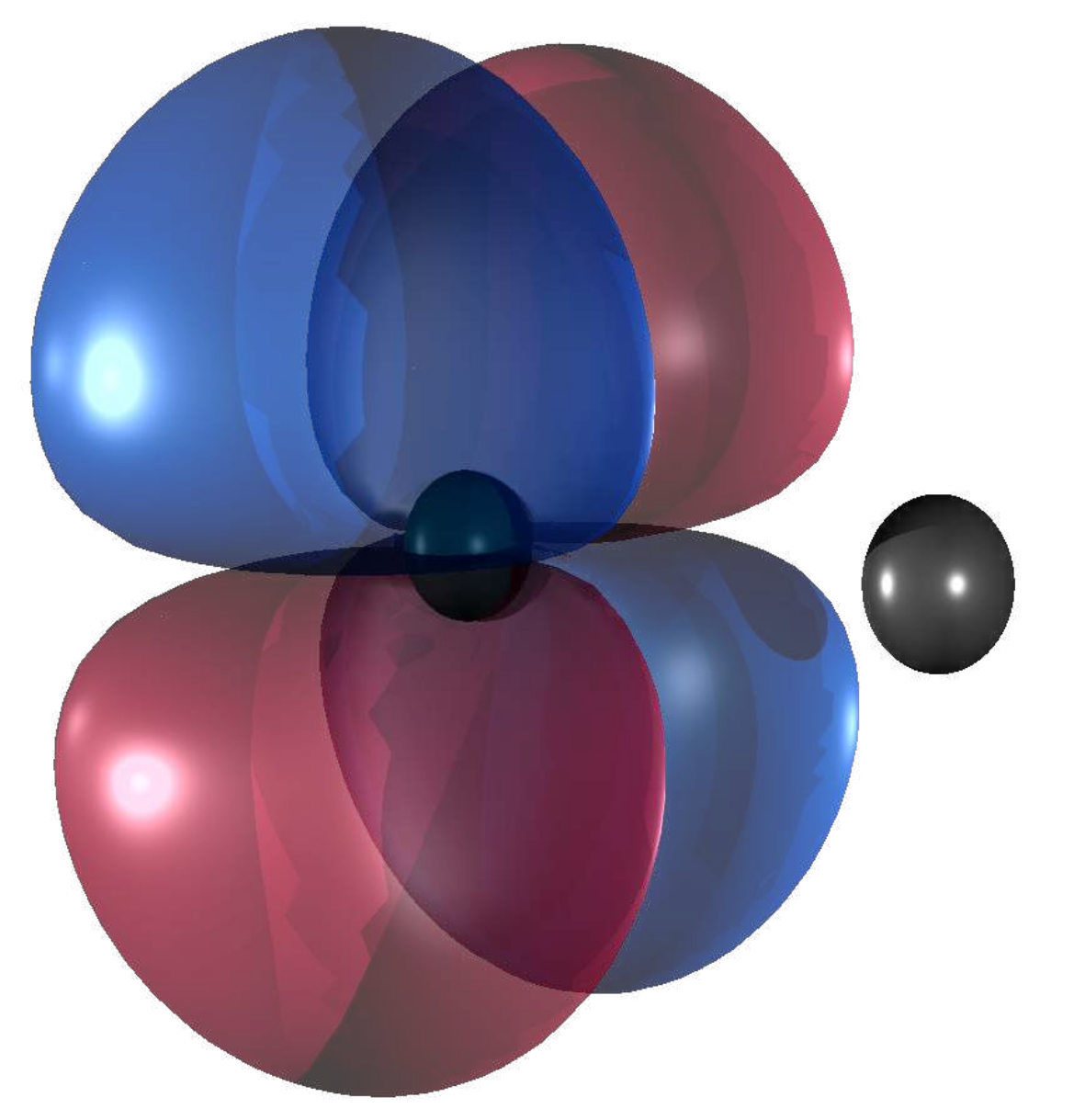}\\[-2pt]
    {\small (a)}
\end{minipage}
\hfill
\begin{minipage}{0.48\columnwidth}
    \centering
    \includegraphics[width=\linewidth]{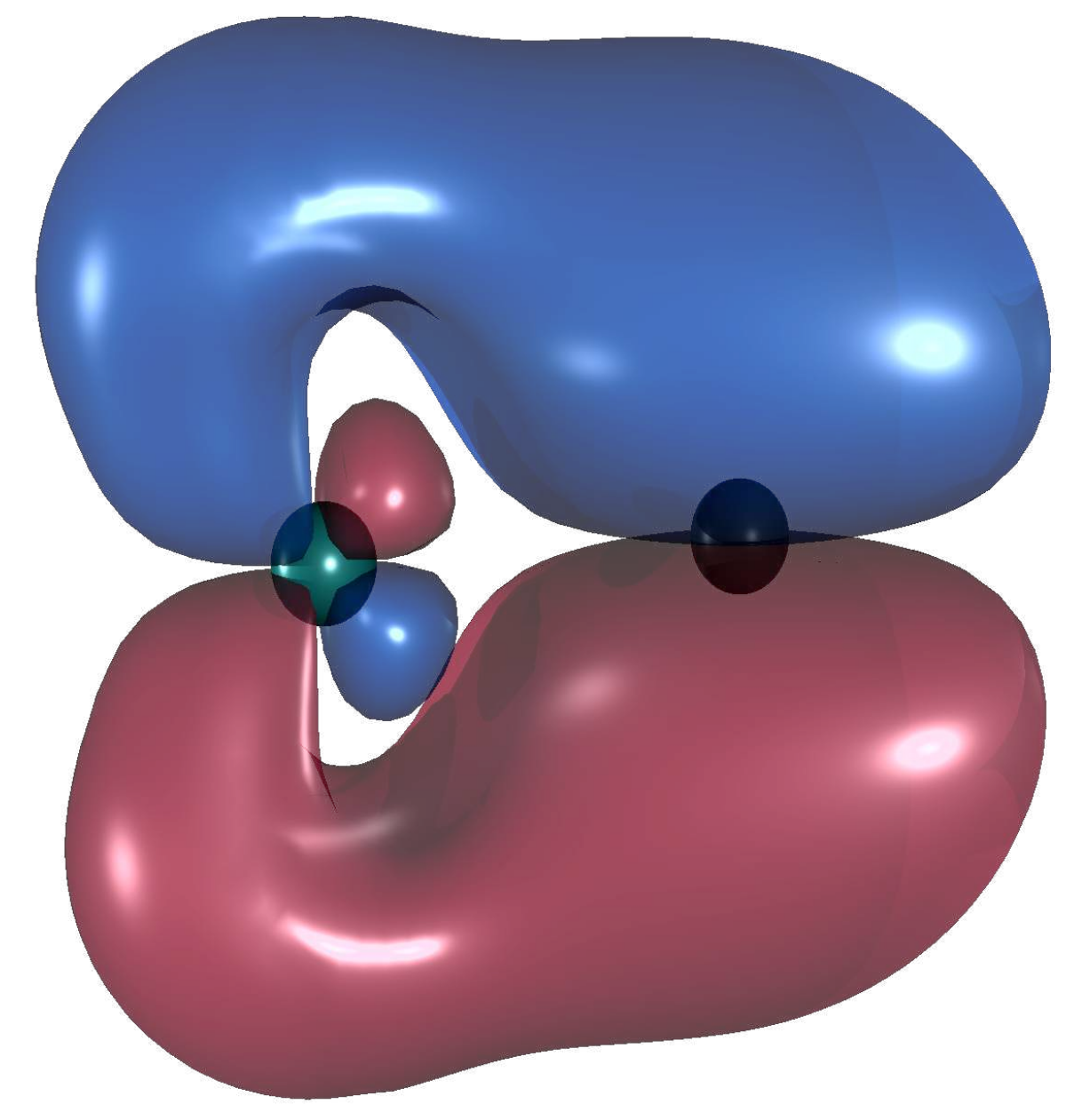}\\[-2pt]
\end{minipage}

\vspace{4pt}

\begin{minipage}{0.48\columnwidth}
    \centering
    \includegraphics[width=\linewidth]{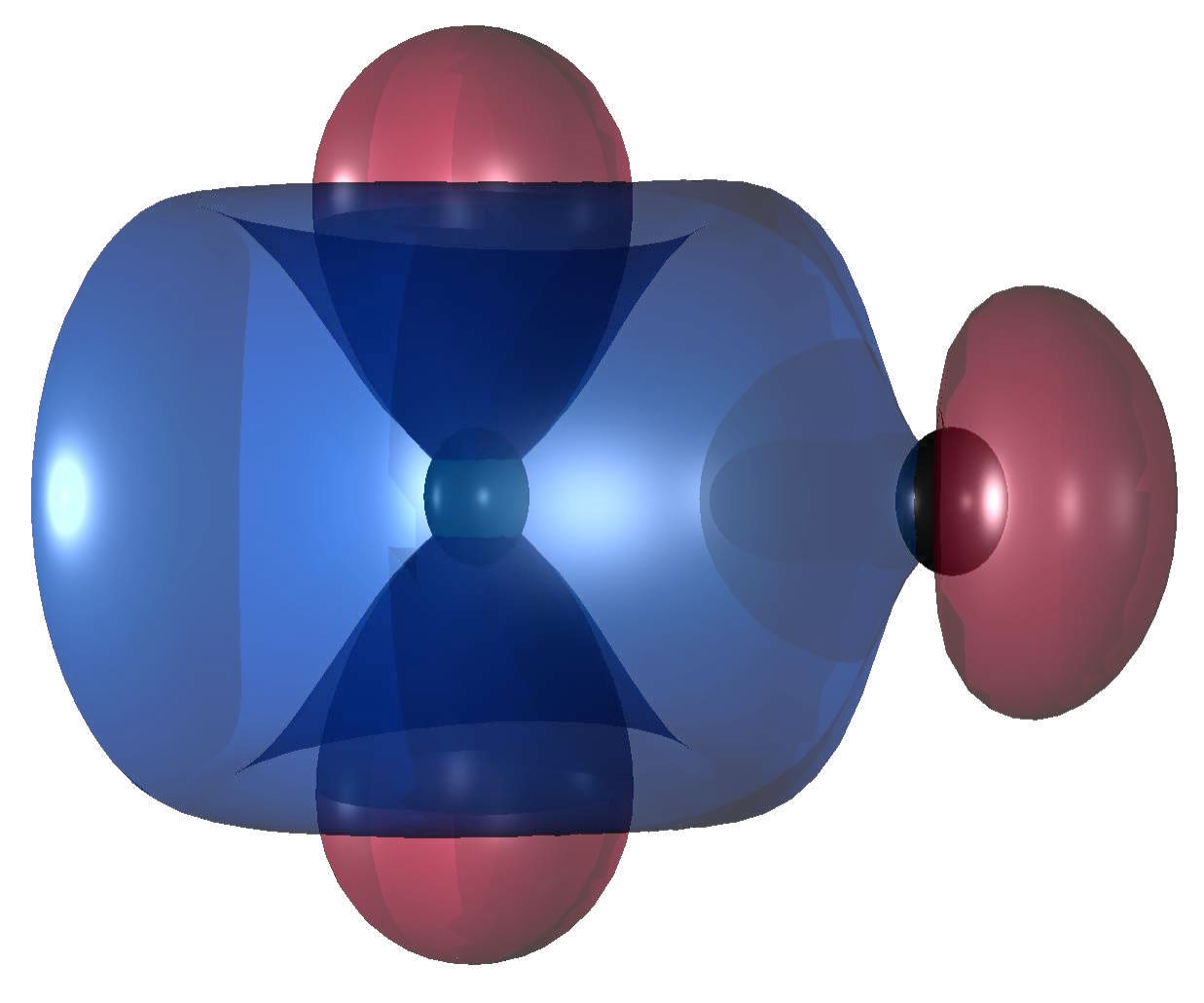}\\[-2pt]
    {\small (b)}
\end{minipage}
\hfill
\begin{minipage}{0.48\columnwidth}
    \centering
    \includegraphics[width=\linewidth]{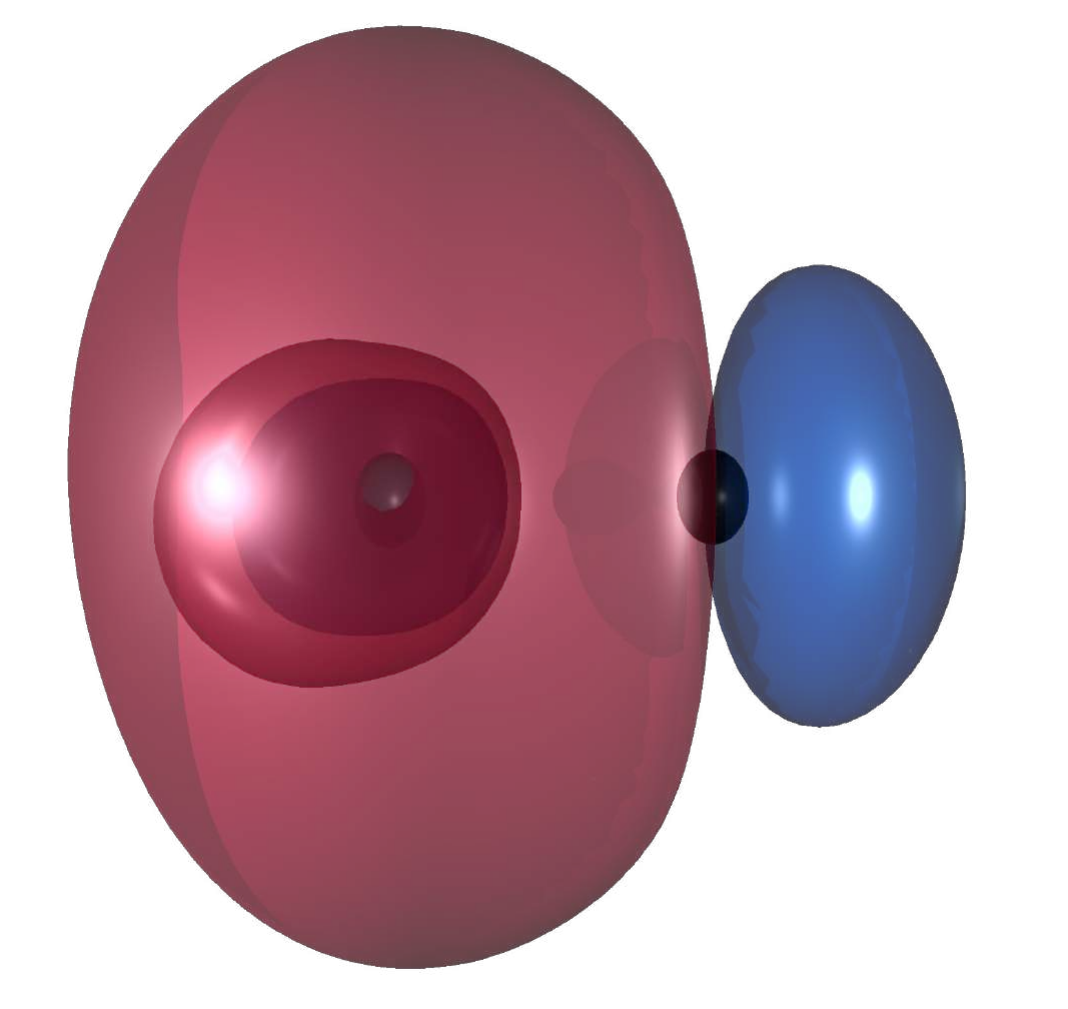}\\[-2pt]
\end{minipage}

\vspace{4pt}

\begin{minipage}{0.48\columnwidth}
    \centering
    \includegraphics[width=\linewidth]{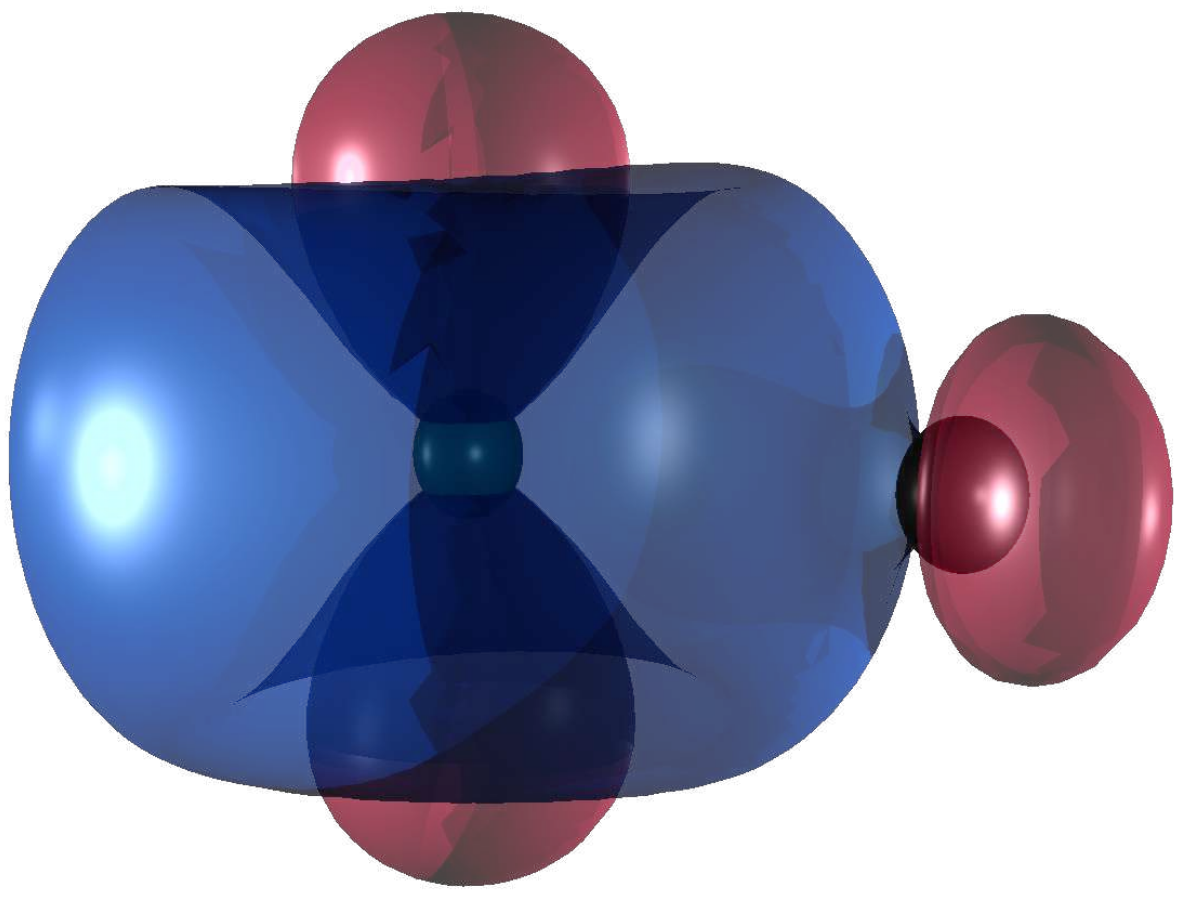}\\[-2pt]
    {\small (c)}
\end{minipage}
\hfill
\begin{minipage}{0.48\columnwidth}
    \centering
    \includegraphics[width=\linewidth]{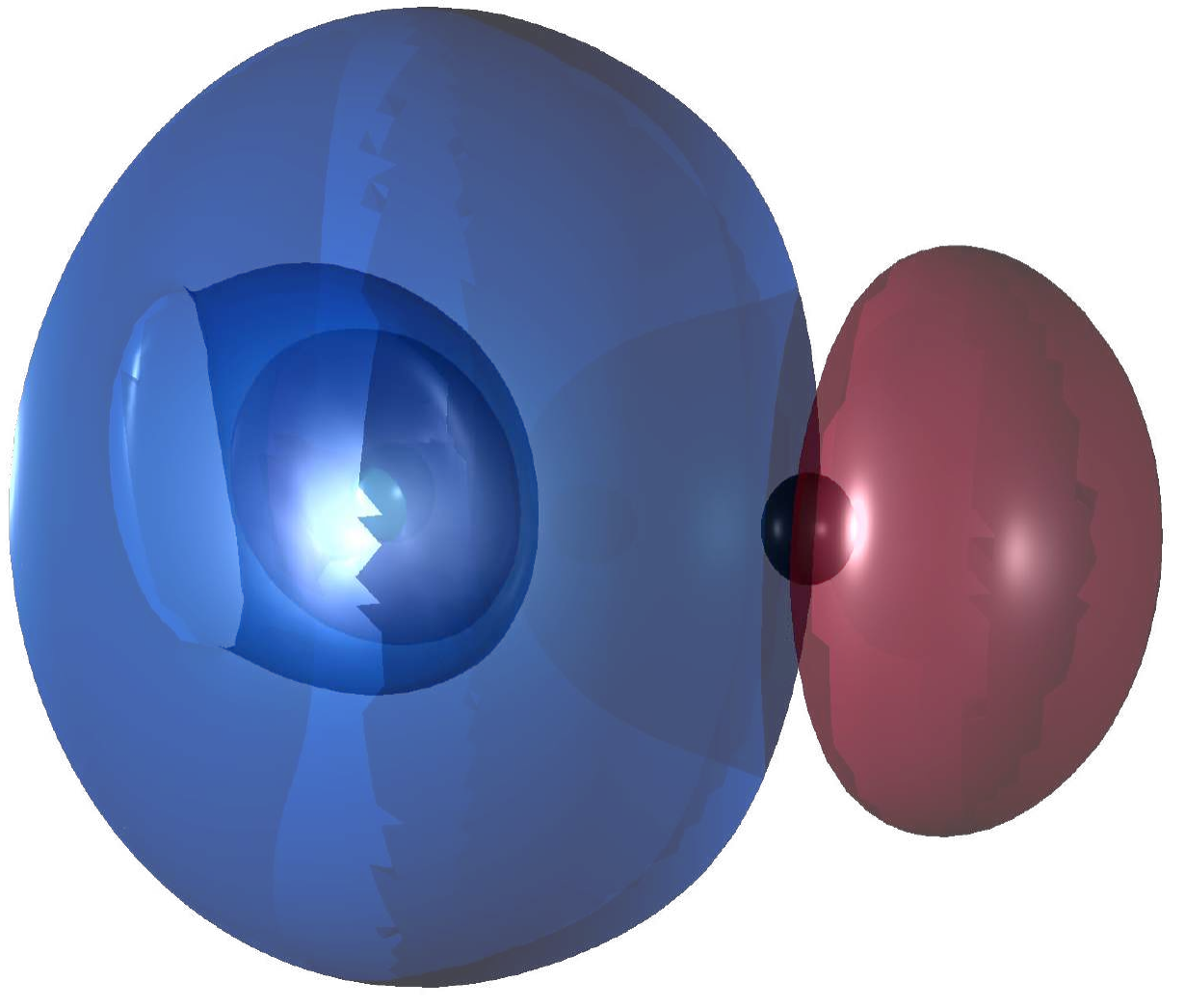}\\[-2pt]
\end{minipage}

\vspace{4pt}

\begin{minipage}{0.48\columnwidth}
    \centering
    \includegraphics[width=\linewidth]{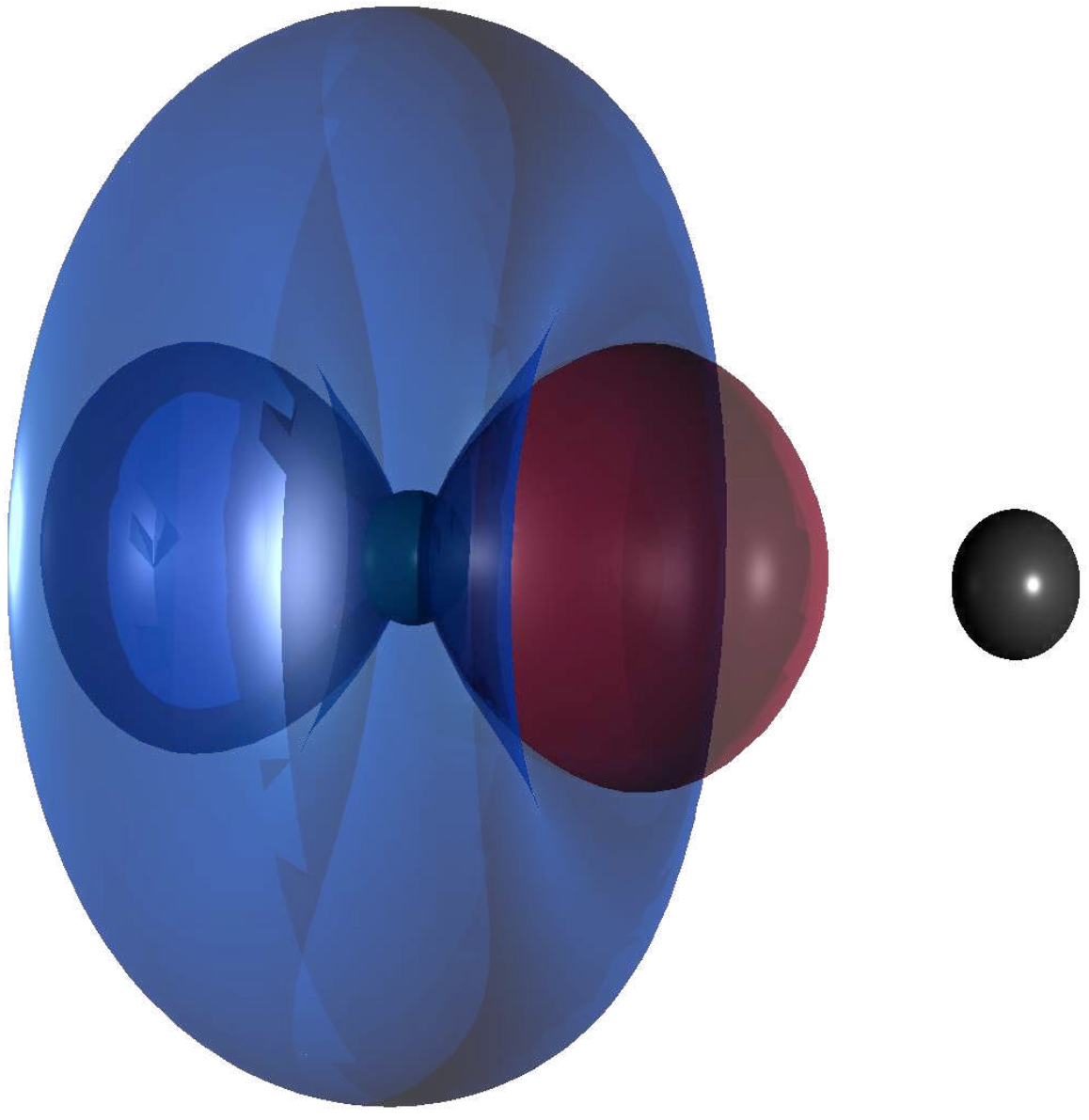}\\[-2pt]
    {\small (d)}
\end{minipage}
\hfill
\begin{minipage}{0.48\columnwidth}
    \centering
    \includegraphics[width=\linewidth]{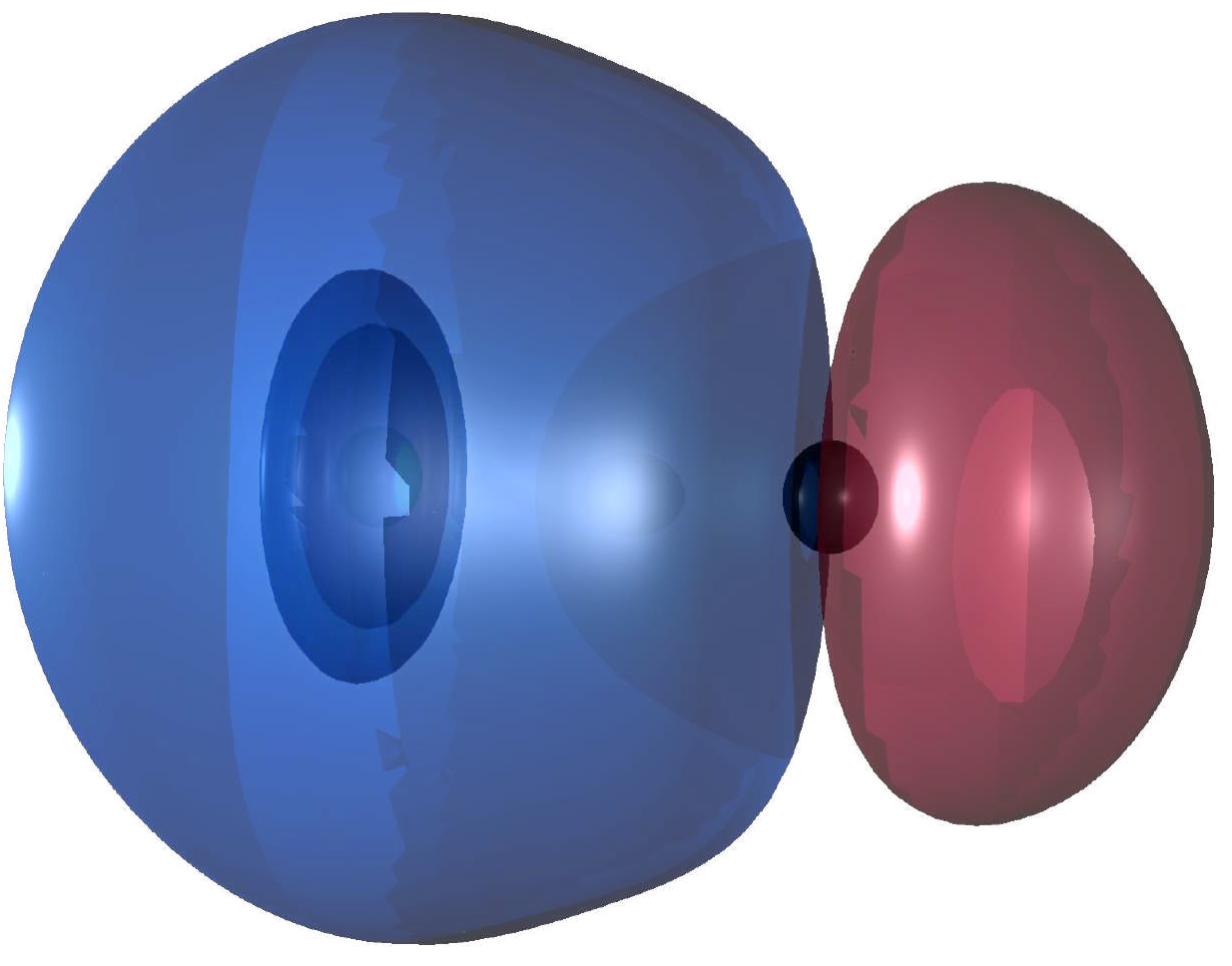}\\[-2pt]
\end{minipage}
\caption{Frontier molecular orbitals of RuC at  $R_{\rm Ru-C}$ = (a) $3.0$~\AA, (b) $3.2$~\AA, (c) $4.0$~\AA, and  (d) $4.2~\text{\AA}$. The HOMO and LUMO are plotted on the left and right panels, respectively. Single-point energies were evaluated using PrTAO-HF(8,8)/aug-cc-pV5Z\cite{Dun89} for C with the corresponding -PP basis for Ru.}
\label{fig:RuC-orbitals}
\end{figure}
The non-monotonic behavior of the Jeffreys divergence in the late-dissociation region of Fig.~\ref{fig:RuC} motivates a closer examination of the underlying Ru--C frontier orbitals. We show in Fig.~\ref{fig:RuC-orbitals} that as the bond is stretched from $R_{\mathrm{Ru-C}}=3.0$ to $3.2$~\AA\ and $R_{\mathrm{Ru-C}}=4.0$ to $4.2$~\AA, the HOMO and LUMO undergo a reorganization, exchanging their dominant spatial character as the corresponding frontier states become nearly degenerate.
The sharp decreases in $J_{\mathrm D}$ coincide with these orbital rearrangements, causing the PrTAO-HF density to become more similar to the CASSCF reference beyond this region. A related nonmonotonic feature is also observed for RHF, indicating that this behavior is not unique to the fractional-occupation construction but instead reflects an underlying reorganization of the Ru--C electronic structure as the system enters the late-dissociation regime. Thus, the maximum in $J_{\mathrm D}$ identifies a region of particularly rapid electronic reorganization, rather than simply a discontinuity introduced by PrTAO-HF.

\section{Computational Details}
All PrTAO-HF and rTAO-HF calculations were performed with a locally modified version of Q-Chem v6.4.\cite{QCHEM5} SCF calculations used a convergence threshold of $10^{-8}$ maximum error and all GHF solutions were checked for internal stability. The Fermi--Dirac occupations were converged self-consistently within the active space according to the protocol in Ref.~\citen{YehYanHsu22} with a convergence threshold of $10^{-8}$ set by the norm of the occupation vector. CCSD and GCCSD calculations were done in PySCF software package,\cite{PySCF1,PySCF2} with an SCF threshold of $10^{-10}$, and a GCCSD amplitude threshold of $10^{-7}$ $\rm E_H$ with $6$ frozen core orbitals for Cr$_3$. Densities were written to cube files with a grid spacing of $0.10$~bohr and $6.0$~bohr margin around the nuclei, except for Cr$_3$, which used $90 \times90 \times 90$ grid with a $2.0$~bohr margin; for each system, all densities were evaluated on the same grid. These cube files were processed using Critic2\cite{critic,critic2} for our $J_{\rm D}$ analysis and each density was normalized to unity. 

\begin{acknowledgments}
The authors thank Alberto Otero-de-la-Roza for productive discussion on GHF spin density comparisons. 
This research was supported in part by the University of Pittsburgh and the University of Pittsburgh Center for Research Computing and Data through the resources provided. Specifically, this work used the H2P cluster, which is supported by NSF award number OAC-2117681.
\end{acknowledgments}

\section*{Data Availability}
All cube file data and data required to reproduce all of the figures in this work are freely available at https://doi.org/10.5281/zenodo.23708148

\section*{Author Contributions}
\noindent\textbf{Kay Carter-Fenk}: conceptualization (lead), supervision (lead), formal analysis (equal), investigation (equal), writing – original draft (equal), writing – review and editing (equal).
\textbf{Md Rafi Ul Azam}: software (lead), formal analysis (equal), investigation (equal), writing – original draft (equal), writing – review and editing (equal).
\textbf{Rachel Campbell}: writing – review and editing (equal), software (supporting), conceptualization (supporting).


%

\end{document}